\documentclass[prb,aps,twocolumn,draft,tightenlines,%
showpacs,floatfix,superscriptaddress]{revtex4}
\usepackage{psfig}
\usepackage{amssymb}
\usepackage{amsmath}
\usepackage{colordvi}

\begin{document}

\title{Spin relaxation time, spin dephasing time and 
ensemble spin dephasing time in $n$-type GaAs quantum wells}

\author{C. L\"u}
\affiliation{Hefei National Laboratory for Physical Sciences at
  Microscale, University of Science and Technology of China, Hefei,
  Anhui, 230026, China}
\affiliation{Department of Physics,
University of Science and Technology of China, Hefei,
  Anhui, 230026, China}
\author{J. L. Cheng}
\affiliation{Department of Physics,
University of Science and Technology of China, Hefei,
  Anhui, 230026, China}
\author{M. W. Wu}
\thanks{Author to whom all correspondence should be addressed.
Mailing address: Department of Physics,
University of Science and Technology of China, Hefei,
  Anhui, 230026, China}
\email{mwwu@ustc.edu.cn.}
\affiliation{Hefei National Laboratory for Physical Sciences at
  Microscale, University of Science and Technology of China, Hefei,
  Anhui, 230026, China}
\affiliation{Department of Physics,
University of Science and Technology of China, Hefei,
  Anhui, 230026, China}
\author{I. C. da Cunha Lima}
\affiliation{Hefei National Laboratory for Physical Sciences at
  Microscale, University of Science and Technology of China, Hefei,
  Anhui, 230026, China}

\date{\today}
\begin{abstract}
  We investigate the spin relaxation and spin dephasing of $n$-type GaAs
  quantum wells. We obtain the
  spin relaxation time $T_1$, the spin dephasing time $T_2$ and the
  ensemble spin dephasing time $T_2^{\ast}$ 
by solving the full microscopic kinetic spin Bloch equations,
 and we show that, analogous
  to the common sense in an isotropic system for conduction electrons,
  $T_1$, $T_2$ and $T_2^{\ast}$ are identical due to the short
  correlation time. The inhomogeneous broadening induced by the
  D'yakonov-Perel  term is suppressed by the scattering, 
especially the Coulomb scattering, in this system.
\end{abstract}

\maketitle

Much attention has been devoted to the spin degree of freedom of
carriers in zinc-blende semiconductors, both in bulk systems and
in reduced dimensionality structures, like quantum wells and
quantum dots. Understanding spin dephasing and spin relaxation of
carriers in these systems is a key factor for the realization of
high quality spintronic devices.\cite{Wolf,Ziese,spintronics,das}
Of  special interest is the calculation of quantities known as
spin relaxation time, $T_1$, and spin dephasing time, $T_2$. $T_1$
is defined as the time it takes for the spins along the
longitudinal field to reach equilibrium. Therefore, it is related
with the relaxation of the average spin polarization. On the other
hand, $T_2$ is defined as the time it takes for the transverse
spins, initially precessing in phase about the longitudinal field,
to lose their phase.\cite{das} In general $T_2 \le 2T_1$, and
$T_1=T_2$ is believed to be true when the system is isotropic and
the correlation time for the interaction is very short compared
with the Larmor period.\cite{Slichter,Bloch}

A qualitative reason for $T_1= T_2$ is that if the correlation
time is short compared with the Larmor period the interaction with
the magnetic fields is not affected by a transformation into a
coordinate system rotating at the Larmor frequency. The
surrounding seems isotropic and the rate of decay will be the same
for all directions. Therefore, longitudinal and transverse
relaxation times will be the same. Hence the decay of the spin
signal will be the same in all directions and $T_1$ equals $ T_2$,
as argued in Ref. \onlinecite{Slichter}. For several years $T_1$
and $T_2$ where considered as the only important factors
describing the spin dynamics under external fields.

In recent years, however, many experiments have been performed
reflecting the dephasing process of the ensemble of electrons,
instead of the dynamics of a single 
one.\cite{kikka} In
fact, electrons with different momentum states have different
precession frequencies due to the momentum dependence of the
effective magnetic field acting on the electron spin, and this
inhomogeneity of precession frequencies can cause a reversible
phase lose. A parameter name, $T_2^\ast$, was coined to describe
the dephasing process associated to this inhomogeneous broadening
of the precessing frequencies.

Wu {\em et al.} have already shown that in the presence of this
inhomogeneous broadening, any scattering, including the
spin-conserving scattering, can cause irreversible spin
dephasing.\cite{wu2,Weng,transport} This
fact leads to the belief that, in general, $T_2^\ast \le T_2$.
However, for conduction electrons $T_2^\ast = T_2$ is known to be
a very good approximation because the inhomogeneous broadening is
always inhibited by the relatively strong scattering existing in
the system.\cite{Dupree}

In this paper, we investigate the spin relaxation and dephasing of
electrons in $n$-type GaAs quantum wells (QWs) grown in the (100)
direction, considered to be the $z$ axis. The width of the well, $a$, is
assumed to be small enough for having just the lowest subband
occupied.  A moderate magnetic field ${\bf B}$ is applied along
the $x$ axis (in the Voigt configuration).

We calculate $T_1$, $T_2$ and $T_2^\ast$ of the electron by
numerically solving the kinetic spin Bloch equations
including scattering by phonons and impurities, besides of the
Coulomb scattering due to electron-electron interaction. Then we
show that $T_1$, $T_2$ and $T_2^\ast$ are identical in these QWs,
as the scattering here is relatively strong.

In the present full microscopic  treatment we associate the above
parameters with the decay slope of the envelope of 
$\rho_{{\bf k},\sigma\sigma^\prime}$, the single-particle density
matrix elements:\\
(i) $T_1$ is determined from the slope of the   envelope of 
\begin{equation}
  \label{sum_N}
\Delta N=\sum_{\bf k}(f_{{\bf k},\uparrow}-f_{{\bf k},\downarrow})\  ;
\end{equation}
(ii) $T_2$ is associated with the
incoherently summed spin coherence\cite{wu1}
\begin{equation}
  \label{sum_rho_incoh}
  \rho= \sum_{\bf k}|\rho_{{\bf k}}(t)| \ ;
\end{equation}
(iii) Finally, $T^{\ast}_2$ is  defined from the slope of the
envelope of the coherently summed spin coherence
\begin{equation}
  \label{sum_rho_coh}
  \rho^{\prime}= |\sum_{\bf k}\rho_{{\bf k}}(t)| \ .
\end{equation}
In these equations  ${\bf \rho}_{{\bf k},\sigma\sigma} \equiv
{f}_{{\bf
    k},\sigma}$ describes the electron distribution functions of
wave vector ${\bf k}$ and spin $\sigma$. The off-diagonal elements
${\rho}_{{\bf k},\uparrow\downarrow} ={\bf \rho}^\ast_{{\bf k},
\downarrow \uparrow}\equiv
{\rho}_{{\bf k}} $ describe the inter-spin-band correlations
for the spin coherence.

With the DP term\cite{dp} included, the Hamiltonian of the electrons can be
written as:
\begin{eqnarray}
H& =&\sum_{{\bf k}\sigma\sigma^\prime}\{\varepsilon_{{\bf
      k}\lambda}\delta_{\sigma\sigma^\prime}
+[g\mu_B{\bf B} + {\bf h}
({\bf k})]\cdot\frac{\mbox{\boldmath$\sigma$\unboldmath}
_{\sigma\sigma^\prime}}{2} \} c^{\dagger}_{{\bf
      k}\sigma}c_{{\bf k}\sigma^{\prime}}\nonumber\\
&&\mbox{}+H_I \ .
\label{total_Hamiltonian}
\end{eqnarray}
Here $\varepsilon_{{\bf k}} =
{\bf k}^2/2m^{\ast}$ is the energy of the electron with wave
vector ${\bf
  k}$. \boldmath$\sigma$\unboldmath\  represents the Pauli matrices.
For wide-band-gap semiconductors such as GaAs, unless a
very large bias voltage is applied,\cite{bias}
 the DP term has its major contribution coming out of the
Dresselhaus term.\cite{dress} Then, we have:
\begin{eqnarray}
  \label{DPx}
   h_x({\bf k}) &=& \gamma k_x(k^2_y - \langle k_z^2\rangle)\ ,\nonumber\\
   h_y({\bf k}) &=& \gamma k_y(\langle k_z^2\rangle - k^2_x)\ ,
   \nonumber\\
   \label{DPy}
   h_z({\bf k}) &=& 0\ .
  \label{DPz}
\end{eqnarray}
Here  $\gamma = (4/3) (m^{\ast} /m_{cv}) (1/\sqrt{2m^{\ast 3}
E_g})(\eta/\sqrt{1-\eta/3})$, $\eta = \Delta/(E_g +\Delta)$ in
which $E_g$ denotes the band gap, $\Delta$ represents the
spin-orbit splitting of the valence band, $m^{\ast}$ stands for
the electron mass in GaAs, and $m_{cv}$ is a constant close in
magnitude to free electron mass $m_0$. In the infinite-well-depth 
approximation, $\langle k_z^2\rangle$ is $(\pi/a)^2$. The
interaction Hamiltonian $H_I$ in Eq.\ (\ref{total_Hamiltonian}) is
composed of the electron-electron Coulomb interaction,
electron-AC-phonon scattering and electron-LO-phonon scattering,
as well as electron-impurity scattering. Their expressions can be
found in textbooks.\cite{Mahan}

We construct the many-body kinetic spin Bloch equations by the
non-equilibrium Green function method\cite{Haug} as follows:
\begin{equation}
  \label{Bloch_eq}
  \dot{\bf \rho}_{{\bf k},\sigma\sigma^\prime} = \dot{\bf \rho}_{{\bf
          k},\sigma\sigma^\prime}|_{coh} +
\dot{\bf \rho}_{{\bf k},\sigma\sigma^\prime}|_{scatt}\ .
\end{equation}
Here  $\dot{\bf \rho}_{{\bf
    k},\sigma\sigma^\prime}|_{coh}$ describes the coherent
spin precessions around the applied magnetic field ${\bf B}$ in
the Voigt configuration, the effective magnetic field ${\bf
h}^({\bf k})$, and the effective magnetic field from the
electron-electron Coulomb interaction in the Hartree-Fock
approximation. This coherent part can be written as:
\begin{widetext}
\begin{eqnarray}
  \label{coh_f}
\left.\frac{\partial f_{{\bf k},\sigma}}{\partial t}\right|_{coh}
& =& - 2 \sigma\{[g\mu_BB + h_x({\bf k})]\mbox{Im}
\rho_{{\bf k}}+h_y({\bf
      k})\mbox{Re}\rho_{{\bf k}}\}
+ 4 \sigma \mbox{Im} \sum\limits_{\bf q} V_{\bf q}
  \rho_{{\bf k}+{\bf q}}^\ast \rho_{{\bf k}} \ ,\\
\left.\frac{\partial \rho_{{\bf k}}}{\partial t}\right|_{coh}  &=&
  \frac{1}{2} [ig\mu_BB + ih_x({\bf k})+h_y
({\bf k})](f_{{\bf k},\uparrow} - f_{{\bf k},\downarrow})
+ i \sum\limits_{\bf q} V_{\bf q} [(f_{{\bf k}+{\bf q},\uparrow} -
  f_{{\bf k},\downarrow}) \rho_{{\bf k}}
- \rho_{{\bf k}+{\bf q}}(f_{{\bf k},\uparrow}- f_{{\bf k}
,\downarrow})]\ ,
\label{coh_rho}
\end{eqnarray}
\end{widetext}
in which $V_{\bf q}$ denotes the Coulomb potential and its expression
can be found in Ref.\ \onlinecite{wu3}.
In Eq.\ (\ref{Bloch_eq}) $\dot{\bf \rho}_{{\bf
    k},\sigma\sigma^\prime}|_{scatt}$
denotes the Coulomb electron-electron, electron-phonon and
electron-impurity scattering. The expressions for these scattering
terms and the details of solving these many-body kinetic spin
Bloch equations are laid out in detail in Ref.\ \onlinecite{wu3}.

\begin{figure}[htb]
  \centerline{
  \psfig{figure=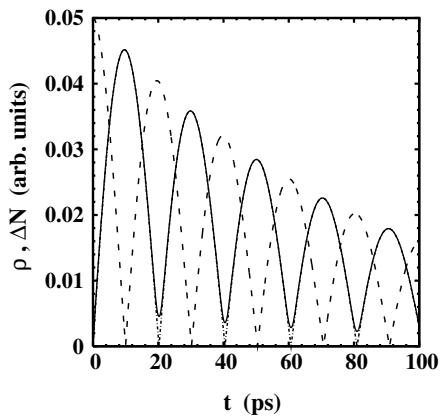,height=5.5cm,angle=0}}
  \caption{Typical time evolution of spin density $\Delta N$ (dotted
curve), the incoherently summed spin coherence $\rho$ (solid
curve) and coherently summed spin coherence $\rho^\prime$ (chained
curve) for the case of $T = 120$\ K, $B=4$\ T, $N=4\times 10^{11}$\ cm$^{-2}$
 and $N_i = 0$.}
\end{figure}

We numerically solve the kinetic spin Bloch equations and obtain
temporal evolution of the electron distribution $f_{{\bf
k},\sigma}(t)$ and the spin coherence $\rho_{{\bf k}}(t)$.  The
material parameters of GaAs in our calculation are the same with
the parameters in Ref.\ \onlinecite{wu3}. In our calculations the
width of the QW is chosen to be 15\ nm; the initial spin
polarization $P_\lambda=\Delta N/N$ is 2.5\ \% and the magnetic field
$B=4$\ T.

In Fig.\ 1 we show the typical evolution of $\rho$,
$\rho^{\prime}$ and $\Delta N$ for $T = 120$ K, the total
electron density $N=4 \times 10^{11}$\ cm$^{-2}$ and
the impurity density $N_i = 0$. It is seen from the figure that
$\rho$, $\rho^{\prime}$ and $\Delta N$ are all oscillating due to
the presence of the magnetic field in Voigt configuration.  
From the envelope of 
$\rho$ and $\Delta N$, we see that $T_1 = T_2$. This result can be
understood as a consequence of the momentum relaxation time here
being less than $1$ picosecond. This is several orders of
magnitude smaller than the period of the effective Larmor
precession induced by the DP term $[2\pi/|{\bf
  h}({\bf k})|]|_{k= k_f}  = 26$\ ps, as well as the Larmor period of magnetic
field $2\pi/\omega_B  = 40$\ ps. Therefore, in the condition of
impurity-free $n$-type GaAs quantum wells, the system is visibly
isotropic in the $x$-$y$--plane and the rate of decay of spin signal
has the same speed in all directions since the correlation time
is short compared with the Larmor period.

We can also observe from Fig.\ 1 that the incoherently summed spin
coherence and the coherently summed spin coherence (solid and chained curves)
are almost identical, which means $T_2 = T_2^{\ast}$. This result
indicates that the inhomogeneous broadening induced by the DP term
is totally suppressed by the strong scattering coming
out of the electron-electron and electron-phonon interactions. To
reveal this effect, we investigate the time evolution of electrons
in different momentum states, which have different precession
frequencies in the presence of the DP term.  This inhomogeneity of
precession frequencies can cause a reversible phase loss making
the coherently summed spin coherence $\rho^\prime$ to decay faster than
the incoherently summed spin coherence $\rho$. However,
this effect can be inhibited by the strong scattering. To
eliminate the effect of the inhibition by strong scattering, we
study the case of $T = 120$\ K, $N=4\times10^{11}$\ cm$^{-2}$ 
and $N_i = 0$ with no Coulomb 
electron-electron scattering included and plot the oscillating
period of each ${\bf k}$ state in Fig.\ 2. In this case, the total
scattering is relatively weak, coming out exclusively from the
scattering by phonons, and we can see that electrons with
different momentum states do have different oscillating periods
although the electron-LO-phonon scattering makes the oscillating
period changing with the period of one LO phonon frequency
as the difference of the diameters of the nearest two concentric circles 
differs exactly by one LO phonon frequency.
However, even in this case, the contribution of the inhomogeneous
broadening is very weak. As discussed in Ref.\ \onlinecite{Allen}
the inhomogeneity of precession frequencies makes a contribution
of $2/\delta\omega_I$ to the total spin dephasing time if the
inhomogeneous lineshape is assumed to be Gaussian. Here
$\delta\omega_I$ represents the root of the mean square of the precession
frequencies and can be written as: $ \delta\omega_I
=\sqrt{\frac{\sum_{\bf k}(\omega_{\bf k} -
   \bar{\omega})^2f_{\bf k}}{\sum_{\bf k} f_{\bf k}} }$,
with $\bar\omega =\frac{\sum_{\bf k}\omega_{\bf k}f_{\bf k}}{\sum_{\bf k}
   f_{\bf k}}$ and $f_{\bf k}$ representing the Fermi
 distribution. In this case the $2/\delta\omega_I$ we calculated is
308\ ps, while the total $T_2$ is only 35\ ps. Therefore, the
contribution of the inhomogeneity can be omitted and the difference
between $T_2$ and $T_2^*$ is still very small.

\begin{figure}[thb]
  \centerline{
  \psfig{figure=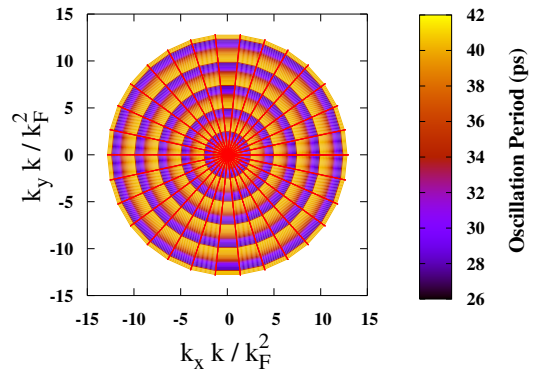,height=6cm,angle=0}}
  \caption{(Color online) The contour plot of oscillating period {\em vs}
 $k_x k/k_F^2$ and  $k_y k/k_F^2$ with $k_F$ representing
 the Fermi wave vector and $k=|{\bf k}|$
for  the case of $T = 120$\ K, $B=4$\ T,
$N=4\times 10^{11}$\ cm$^{-2}$  and $N_i = 0$. There is no 
electron-electron Coulomb scattering in the calculation.}
\end{figure}

Furthermore, when we include the Coulomb scattering in this system, 
we find that electrons in each momentum state has the
{\em same} oscillating period of 40\ ps, which is exactly equals to the
Larmor period induced by the magnetic field and also the
oscillating period of $\rho$ and $\rho^{\prime}$. The
inhomogeneous broadening is suppressed and $T_2$ equals
$T_2^{\ast}$.

We further check this result with different temperatures,
electron densities and impurity densities. We find that
 $T_1 = T_2 = T_2^\ast$ is valid in a very wide range of
temperature from 10\ K to 300\ K, and  electron densities from
$2\times 10^{10}$ to $4\times 10^{11}$\ cm$^{-2}$. Including the
impurity scattering will not change this result. Even in the case
of $T = 10$\ K,  with the total electron density $N=2\times 
10^{10}$\ cm$^{-2}$ and without the impurity and the Coulomb
scattering, where the electron-AC-phonon scattering is the
dominant scattering, the difference obtained between $T_1$, $T_2$
and $T_2^\ast$ is still less than 6\ \%.

In conclusion, we have investigated the spin relaxation and the spin
dephasing of electrons in $n$-type GaAs quantum wells and
calculate $T_1$, $T_2$ and $T_2^\ast$ by numerically solving the
kinetic spin Bloch equations. We have obtained that they
have the same value in a very wide range of temperatures, electron
densities and the impurity densities and we have shown that this
behavior is due to the short correlation time. More experiments such as
the spin echo experiment\cite{echo} are needed to check the findings 
here.

This work was supported by the Natural Science Foundation of China
under Grant No.\ 10574120,  the National Basic Research Program of
China under Grant No.\ 2006CNBOL1205, the
Knowledge Innovation Project of Chinese Academy of Sciences and
SRFDP. I.C.C.L. was partially supported by CNPq from Brazil.
 One of the authors (C.L.) would like to thank J. Zhou for
providing the code of electron-AC phonon scattering.


\begin{thebibliography}{0}
\bibitem{Wolf} S. A. Wolf, J. Supercond. {\bf 13}, 195 (2000).
\bibitem{Ziese} {\em Spin Electronics}, edited by M. Ziese and
  M. J. Thornton (Springer, Berlin, 2001).
\bibitem{spintronics} {\it Semiconductor Spintronics and Quantum
 Computation}, ed. by D. D. Awschalom, D. Loss, and N. Samarth
    (Springer-Verlag, Berlin, 2002).
\bibitem{das} I. \v Zuti\'c, J. Fabian, and S. Das Sarma,
  Rev. Mod. Phys. {\bf 76}, 323 (2004).
\bibitem{Slichter} D. Pines and C. P. Slichter, Phys. Rev. {\bf 100},
  1014 (1955).
\bibitem{Bloch} R. K. Wangsness and F. Bloch,
  Phys. Rev. {\bf 89},  728 (1953).
\bibitem{kikka} J. M. Kikkawa, I. P. Smorchkova, N. Samarth, 
and D. D. Awschalom, Science {\bf 277}, 1284
  (1997); J. M. Kikkawa and D. D. Awschalom, Phys. 
Rev. Lett. {\bf 80}, 4313 (1998); 
Science {\bf 281}, 656 (2000);  Science  {\bf 287}, 473 (2000);
G. Salis, D. T. Fuchs, J. M. Kikkawa, D. D. Awschalom, Y. Ohno and H. Ohno,
Phys. Rev. Lett. {\bf 86}, 2677 (2001); 
Y. Kato, R. C. Myers, D. C. Driscoll, A. C. Gossard, J. Levy, and 
D. D. Awschalom, Science {\bf 299}, 1201 (2003); H. Hoffmann, 
G. V. Astakhov, T. Kiessling, W. Ossau, G. Karczewski, T. Wojtowicz, 
J. Kossut, and L. W. Molenkamp, Phys. Rev.
B {\bf 74}, 073407 (2006).
\bibitem{wu2} M. W. Wu and C. Z. Ning, Eur. Phys. J. B {\bf 18},
  373 (2000); M. W. Wu, J. Phys. Soc. Jpn. {\bf 70}, 2195 (2001);
J. L. Cheng, M. Q. Weng, and M. W. Wu,  Solid State Commun. 
{\bf 128}, 365 (2003).
\bibitem{Weng} M. Q. Weng and M. W. Wu, Phys. Rev. B {\bf 68},
  075312 (2003); {\bf 71}, 199902(E) (2005);  Chin. Phys. Lett.
{\bf 22}, 671 (2005);  M. Q. Weng and M. W. Wu, 
Phys. Rev. B {\bf 70}, 195318 (2004);
L. Jiang and M. W. Wu, {\em ibid.} {\bf 72}, 033311 (2005);
C. L\"u, J. L. Cheng, and M. W. Wu, {\em ibid.} {\bf 73}, 125314 (2006). 
\bibitem{wu3} M. Q. Weng, M. W. Wu, and L. Jiang, Phys. Rev. B {\bf
    69}, 245320 (2004); J. Zhou, J. L. Cheng, and M. W. Wu, {\em ibid.}
{\bf 75}, 045305 (2007).
\bibitem{transport}M. Q. Weng and M. W. Wu, J. Appl. Phys. {\bf 93},
410 (2003);  M. Q. Weng, M. W. Wu,
and Q. W. Shi, Phys. Rev. B {\bf 69}, 125310 (2004); L. Jiang, M. Q. Weng,
 M. W. Wu, and J. L. Cheng, J. Appl. Phys. {\bf 98}, 113702 (2005).
\bibitem{wu1}M. W. Wu and H. Metiu, Phys. Rev. B {\bf 61}, 2945 (2000).
\bibitem{Dupree} R. Dupree and B. W. Holland, Phys. Stat. Sol. (b) {\bf
    24}, 275 (1967).
\bibitem{dp}  M. I. D'yakonov and V. I.  Perel',  
Zh. Eksp. Teor. Fiz. {\bf 60}, 1954 (1971) 
[Sov. Phys. JETP {\bf 38},  1053 (1971)].
\bibitem{bias} W. H. Lau and M. E. Flatt\'e, Phys. Rev. B {\bf 72}, 
161311(R) (2005).
\bibitem{dress} G. Dresselhaus, Phys. Rev. {\bf 100}, 580 (1955).
\bibitem{Mahan} G. D. Mahan, {\em Many-particle Physics} (Plenum, New
  York, 1981).
\bibitem{Haug} H. Haug and A. P. Jauho, {\em Quantum Kinetics in
    Transport and Qptics of Semiconductors} (Springer-Verlag, Berlin,
  1996).
\bibitem{Allen} L. Allen and J. H. Eberly, {\em Optical Resonance and
    Two-level Atoms} (Dover, New York, 1975).
\bibitem{echo} A. M. Tyryshkin, S. A. Lyon, W. Jantsch, and F. Sch\"affler,
Phys. Rev. Lett. {\bf 94}, 126802 (2005).

\end{thebibliography}
\end{document}